\documentclass[11pt]{article}
\usepackage[a4paper, total={6.5in, 10in}]{geometry}
\usepackage{eqnarray,amsmath}
\usepackage{hyperref,url}
\usepackage{mathtools }
\usepackage{enumerate}
\usepackage{framed,lipsum}
\usepackage{tikz}
\usepackage{tikz-network}
\usetikzlibrary{shapes}
\usepackage{float}
\usepackage{amssymb}
\usepackage{amsthm}
\usepackage{bm}
\usepackage{cases}
\usepackage{authblk}

\newtheorem{theorem}{Theorem}
\newtheorem{lemma}[theorem]{Lemma}

\newtheorem{observation}[theorem]{Observation}
\newtheorem{fact}[theorem]{Fact}

\newtheorem{definition}{Definition}

\newcommand{\etal}{{\em et al. }}

\newcommand{\COMMENTED}[1]{{}}

\newcommand{\cost}{\ensuremath{\text{\footnotesize\textsf{COST}}}}
\newcommand{\rev}{\ensuremath{\text{\footnotesize\textsf{REV}}}}
\newcommand{\mergecost}{\ensuremath{\text{\footnotesize\textsf{MERGE-COST}}}}
\newcommand{\mergerev}{\ensuremath{\text{\footnotesize\textsf{MERGE-REV}}}}

\title{ Hierarchical Clustering via Local Search}

\author{‌Hossein Jowhari}

\affil{Faculty of Mathematics, \\ K. N. Toosi University of Technology, Tehran, Iran \footnote{Email: jowhari@kntu.ac.ir}}

\begin{document} 
\maketitle

\begin{abstract}
In this paper, we introduce a local search algorithm for hierarchical clustering. For the local step, we consider a tree re-arrangement operation, known as the {\em interchange}, which involves swapping two closely positioned sub-trees within a tree hierarchy. The interchange operation has been previously used in the context of phylogenetic trees. As the objective function for evaluating the resulting hierarchies, we utilize the revenue function proposed by Moseley and Wang (NIPS 2017.) 

In our main result, we show that any locally optimal tree guarantees a revenue of at least $\frac{n-2}{3}\sum_{i < j}w(i,j)$ where is $n$ the number of objects and $w: [n] \times [n] \rightarrow \mathbb{R}^+$ is the associated similarity function. This finding echoes the previously established bound for the average link algorithm as analyzed by Moseley and Wang. We demonstrate that this alignment is not coincidental, as the average link trees enjoy the property of being locally optimal with respect to the interchange operation. Consequently, our study provides an alternative insight into the average link algorithm and reveals the existence of a broader range of hierarchies with relatively high revenue achievable through a straightforward local search algorithm.

Furthermore, we present an implementation of the local search framework, where each local step requires $O(n)$ computation time. Our empirical results indicate that the proposed method, used as post-processing step, can effectively generate a hierarchical clustering with substantial revenue.
\end{abstract}

\section{Introduction} 
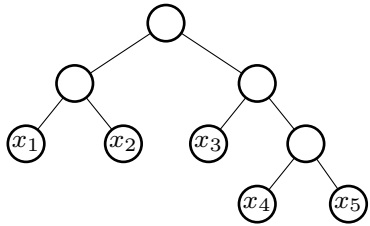
\begin{figure}[H]
\label{fig:HCtree}
\centering
\begin{tikzpicture}[scale=0.8,transform shape]

\Vertex[x=5,y=5, size = 0.6,  color=white, fontscale=1.5]{x1}

\Vertex[x=3.5, y=4, size = 0.6,  color=white, fontscale=1.5]{x2}
\Vertex[x=6.5,y=4, size = 0.6,  color=white, fontscale=1.5]{x3}

\Vertex[x=2.7, y=3, size = 0.6, label=$x_1$, color=white, fontscale=1.5]{x4}
\Vertex[x=4.3,y=3, size = 0.6, label=$x_2$, color=white, fontscale=1.5]{x5}
\Vertex[x=5.7, y=3, size = 0.6, label=$x_3$, color=white, fontscale=1.5]{x6}
\Vertex[x=7.3,y=3, size = 0.6, color=white, fontscale=1.5]{x7}

\Vertex[x=6.5, y=2, size = 0.6, label=$x_4$, color=white, fontscale=1.5]{x8}
\Vertex[x=8,y=2, size = 0.6, label=$x_5$, color=white, fontscale=1.5]{x9}


\draw (x2) -- (x1);
\draw (x3) -- (x1);
\draw (x4) -- (x2);
\draw (x5) -- (x2);
\draw (x6) -- (x3);
\draw (x7) -- (x3);
\draw (x8) -- (x7);
\draw (x9) -- (x7);

\end{tikzpicture}
\caption{A HC tree over a dataset containing five data points.}
\end{figure}

Hierarchical clustering is a technique utilized in data mining and exploratory data analysis to group similar data points into clusters. In contrast to other clustering methods, hierarchical clustering organizes data into a tree-like hierarchy of sub-clusters, referred to as a dendrogram or a HC tree. In the hierarchy, the root of the tree represents all data points while each internal node defines a subcluster. The individual data points are situated at the leaves of the tree. Figure \ref{fig:HCtree} illustrates an HC tree over a dataset containing five elements.

Two build a hierarchical clustering, two main approaches are commonly used: agglomerative and divisive. The agglomerative approach starts by assigning each data point to its own cluster, then merges clusters until only one cluster remains. Different distance or similarity measures are used to decide which clusters to merge. On the other hand, the divisive approach begins with all data points in one cluster, which is then divided into smaller clusters recursively based on some split strategy. We refer the reader to the survey \cite{MurtaghC12} for an overview of the classical algorithms in hierarchical clustering.

\subsection{Hierarchical clustering as an optimization problem}
There are various unsupervised objective functions for evaluating a hierarchical clustering. In this work, we consider Dasgupta's cost funtion  \cite{Dasgupta16} and its dual variant as studied by Moseley and Wang \cite{MoseleyW17}. Here the input data consists of a symmetric function $w: [n] \times [n] \rightarrow \mathbb{R}^+$ which specifies the pairwise similarity between $n$ data points; thus $w(i,j)$ denotes the similarity between the objects $i$ and $j$. Given a hierarchical clustering represented as a HC tree $T$ over $n$ leaves, where each leaf is labeled uniquely by a number in $[n]=\{1,\cdots, n\}$, the clustering quality of $T$ is evaluated as follows.  For the leaves $i$ and $j$, let  $T_{i,j}$ denote the subtree rooted at the lowest common ancestor of $i$ and $j$. Let
$|T_{i,j}|$ denote the number of leaves in $T_{i,j}$. Dasgupta's cost function is defined as follows. 

\begin{equation}
\label{def:dasgupta}
\text{cost}(T) = \sum_{i < j} w(i,j) |T_{i,j}|.
\end{equation}

Moseley and Wang's revenue function as stated below is the dual version of Dasgupta's cost function.

\begin{equation}
\label{def:MW}
\text{rev}(T) = \sum_{i < j} w(i,j) (n-|T_{i,j}|)
\end{equation}

Note that, intuitively, when the similarity between the pair of objects $i$
and $j$ is high, maximizing the revenue function (or equivalently minimizing the cost function) tends to place the pair closer together, ideally within the leaves of a small subtree. In \cite{Dasgupta16}, it was noted that an optimal solution under his formulation has the structure of a binary tree, and moreover computing the optimal tree is a NP-hard problem. Same holds for the dual variant of the problem. As a result, the endeavor to identify a tree with minimal cost (or maximum revenue) has been approached from the perspective of approximation algorithms \cite{Dasgupta16, MoseleyW17, CharikarC17, Cohen-AddadKMM18}. There is also a line of work that analyses the performance of classical algorithms for hierarchical clustering through the lens of Dasgupta's cost function and its variants. In a particular case,  
 Moseley and Wang \cite{MoseleyW17} have established the following fact regarding the popular agglomerative HC algorithm known as the average link. 

\begin{fact}
\label{fact:MWavglink}
 \cite{MoseleyW17} Let $T$ be the result of the average link algorithm. Then
$\rev(T)  \ge \frac{n-2}3\sum_{i < j} w(i,j)$. 
\end{fact}
 
Since $(n-2)\sum_{i < j} w(i,j)$ serves as an upper bound for the optimal revenue, it follows that the average link generates trees yielding revenues of at least $1/3$ of the optimal revenue. It's worth noting that polynomial-time algorithms with improved approximation factors exist \cite{CharikarCN19, ChatziafratisYL20, AlonAV20}, although these findings rely on semidefinite programming.

\subsection{Local search for hierarchical clustering}
A common approach for solving an optimization problem is by alliteratively enhancing an initial solution through a series of local adjustments. These adjustments continue until a point is reached where further local modifications no longer enhance the quality of the solution, leading to a local optimum. In the case of clustering problems, the local search technique has demonstrated success across various non-hierarchical settings (see the related works section for some pointers.) In this study, we investigate a natural local search heuristic tailored for hierarchical clustering. Before we express our local search algorithm, we give a formal definition of a HC tree as used in this paper. 
%

\begin{definition}  A HC tree of order $n$ is a full binary tree \footnote{A full binary tree is a rooted tree where each non-leaf has two children.} with $n$ leaves where each leaf is labeled uniquely by a number in the set $[n]$. There are no specific ordering for the children in a HC tree. 
\end{definition}

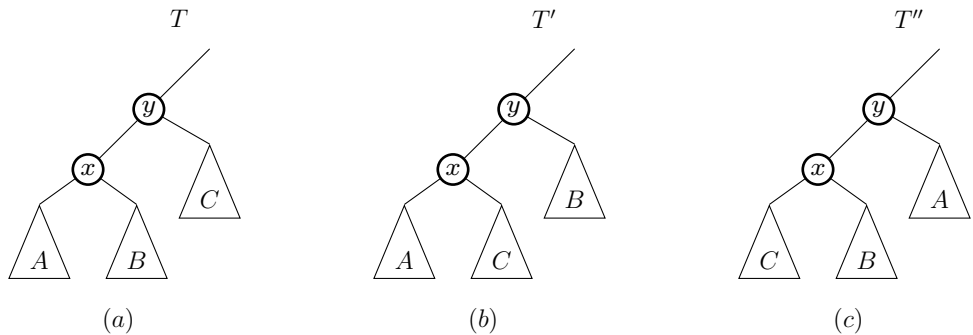
\begin{figure}[H]
\centering
\begin{tikzpicture}[scale=0.8,transform shape]

\Vertex[x=2.5,y=5.5, size = 0.5, label=$x$, color=white, fontscale=1.5]{x1}
\Vertex[x=3.5, y=6.5, size = 0.5, label=$y$, color=white, fontscale=1.5]{y1}

\node[isosceles triangle,
    draw,
    rotate=90,
    fill=white,
    minimum size =1cm] (A1) at (1.7,4){};

\Text[x=1.7, y =4]{\large $A$}

\node[isosceles triangle,
    draw,
    rotate=90,
    fill=white,
    minimum size =1cm] (B1) at (3.3,4){};

\Text[x=3.3, y = 4]{\large $B$}

\node[isosceles triangle,
    draw,
    rotate=90,
    fill=white,
    minimum size =1cm] (C1) at (4.5,5){};
    
\Text[x=4.5, y = 5]{\large $C$}
  
\draw (x1) -- (A1.apex);
\draw (y1) -- (x1);
\draw (x1) -- (B1.apex);
\draw (y1) -- (C1.apex);
\draw (y1) -- (4.5,7.5);

\Text[x=4, y = 8]{\large $T$}

\Text[x=3, y = 3]{\large $(a)$}


\Vertex[x=8.5,y=5.5, size = 0.5, label=$x$, color=white, fontscale=1.5]{x2}
\Vertex[x=9.5, y=6.5, size = 0.5, label=$y$, color=white, fontscale=1.5]{y2}

\node[isosceles triangle,
    draw,
    rotate=90,
    fill=white,
    minimum size =1cm] (A2) at (7.7,4){};

\Text[x=7.7, y =4]{\large $A$}

\node[isosceles triangle,
    draw,
    rotate=90,
    fill=white,
    minimum size =1cm] (B2) at (9.3,4){};

\Text[x=9.3, y = 4]{\large $C$}

\node[isosceles triangle,
    draw,
    rotate=90,
    fill=white,
    minimum size =1cm] (C2) at (10.5,5){};
    
\Text[x=10.5, y = 5]{\large $B$}
  
\draw (x2) -- (A2.apex);
\draw (y2) -- (x2);
\draw (x2) -- (B2.apex);
\draw (y2) -- (C2.apex);
\draw (y2) -- (10.5,7.5);

\Text[x=10, y = 8]{\large $T'$}
\Text[x=9, y = 3]{\large $(b)$}

\Vertex[x=14.5,y=5.5, size = 0.5, label=$x$, color=white, fontscale=1.5]{x2}
\Vertex[x=15.5, y=6.5, size = 0.5, label=$y$, color=white, fontscale=1.5]{y2}

\node[isosceles triangle,
    draw,
    rotate=90,
    fill=white,
    minimum size =1cm] (A2) at (13.7,4){};

\Text[x=13.7, y =4]{\large $C$}

\node[isosceles triangle,
    draw,
    rotate=90,
    fill=white,
    minimum size =1cm] (B2) at (15.3,4){};

\Text[x=15.3, y = 4]{\large $B$}

\node[isosceles triangle,
    draw,
    rotate=90,
    fill=white,
    minimum size =1cm] (C2) at (16.5,5){};
    
\Text[x=16.5, y = 5]{\large $A$}
  
\draw (x2) -- (A2.apex);
\draw (y2) -- (x2);
\draw (x2) -- (B2.apex);
\draw (y2) -- (C2.apex);
\draw (y2) -- (16.5,7.5);

\Text[x=16, y = 8]{\large $T''$}
\Text[x=15, y = 3]{\large $(c)$}
\end{tikzpicture}
\caption{An interchange operation performed on the edge $(x,y)$. }
\label{fig:interchange}

\end{figure}

Our local adjustment, defined for HC trees, is based on a transformation known as the {\em nearest neighbor interchange}, or simply the {\em interchange}.   Initially introduced in the context of evolutionary biological trees by Moore \etal \cite{MGBarnabas73}, this operation \footnote{The interchange operation has been originally defined for a specific family of un-rooted trees (see Figure 1 in the reference \cite{LiTZ96}) while here we adopt it for rooted binary trees.} has been further analyzed in subsequent works  \cite{WSmith78,CulikW82,LiTZ96, DasGuptaH0TZ16}. The interchange operation applies to an edge $(x,y)$ where $x$ is a non-leaf and $y$ is the parent of $x$, as shown in the Figure \ref{fig:interchange}(a). It yields one of the outcomes (b) or (c) in the same figure.

We emphasize the fact that our HC trees are un-ordered. Consequently, the node $x$ in Figure \ref{fig:interchange} can be situated at both left or right child of the parent node $y$ if we imagine an ordering on the tree. Now, given the definition of the interchange operation and Dasgupta's cost function, we can establish the concept of a locally optimal tree as follows.

\begin{definition}
\label{def:locallyoptimal}
Given a HC tree $T$, we say an interchange operation is profitable if it increases $\rev(T)$, or equivalently if it decreases $\cost(T)$.  If no interchange is profitable for a given tree $T$, we classify $T$ as locally optimal.
\end{definition}

\paragraph{Our contributions.} In our main result, we demonstrate that any locally optimal tree guarantees a revenue of at least $\frac{n-2}3\sum_{i < j} w(i,j)$.  This finding aligns with the bound previously established for the average link trees in \cite{MoseleyW17} as stated in Fact \ref{fact:MWavglink}. Interestingly, our analysis reveals that average link trees indeed fall within the category of locally optimal trees, hence providing an alternative proof of Fact \ref{fact:MWavglink}. In comparison with the analysis in \cite{MoseleyW17}, establishing the bound for the locally optimal trees (which is a more general class of trees in comparison with average link trees) turns out to be more difficult. Fortunately, through an inductive argument and by utilizing the chain of local relationships in the tree, we are able to derive a global bound for the whole structure.   Here we should mention that a random binary tree also achieves a revenue of $\frac{n-2}3\sum_{i < j} w(i,j)$ in expectation as shown in \cite{MoseleyW17}. However a random tree is far from being locally optimal as our experiments suggest. Consequently, our work offers a theoretical underpinning for the observed practical effectiveness of the average link heuristic, which consistently yields trees with significantly higher revenue compared to the lower bound of $\frac{n-2}3\sum_{i < j} w(i,j)$ as confirmed by our experiments and previous studies \cite{MenonRSCCK19}.

Our local search framework also suggests a metric distance function over HC trees. Such a distance function can be useful in comparing hierarchies generated by different methods. We offer insights that shows any HC tree of order $n$ can be converted into another HC tree of order $n$ using at most $O(n\log n)$ number of interchange operations.  

Furthermore, we give a straightforward implementation of the local search method utilizing elementary data structures. Our implementation leverages an array-based data structure with space complexity $O(n^2)$ while each local improvement is executed in $O(n)$ time. 

Finally in our empirical results, we examine the performance of two variants of the local search method over real datasets. Our experiments suggest that while initiating the method from a random tree can take a long time time to coverage, the local search method can be effective as a post-processing step to boost the quality of the solutions generated by other methods. 

\subsection{Related Works}
Local search algorithms for clustering problems (in the non-hierarchical settings) have been the subject of numerous studies  \cite{AryaGKMMP04,KanungoMNPSW04,FriggstadRS19,Cohen-AddadKM19,Cohen-AddadGHOS22}. In particular, the work by Arya \etal \cite{AryaGKMMP04} analyses a swapping-based local search algorithm for $k$-median objective function. Recent studies \cite{LattanziS19,ChooGPR20, BCLP23} have shown that a local search phase followed by the popular $k$-means++ algorithm can yield constant factor approximations for the $k$-means problem.

We are not aware of previous works on local search for hierarchical clustering except possibly an algorithm by Moseley and Wang \cite{MoseleyW17}. The result in \cite{MoseleyW17} is a divisive top-down algorithm where local search is used as a split strategy for dividing the subclusters.  Hence their algorithm is not a local search algorithm in the usual sense when one begins from an initial full-blown solution and applies a series of local adjustments to reach a local optimum. Additionally, the adjustment operation used in our study has been previously  employed as a tree re-arrangement strategy in the works \etal \cite{KobrenMKM17, MonathKKGM19} within the context of incremental algorithms for hierarchical clustering. The rotate operation, as employed in \cite{KobrenMKM17, MonathKKGM19} is in fact identical to the interchange operation used in our work. However these works do not utilize the adjustment as a local search strategy, and furthermore the quality of the hierarchical clustering is evaluated based on dendrogram purity \cite{HellerG05} which is a supervised way of assessing a hierarchical clustering.

The interchange operation, initially termed as {\em nearest neighbor 1-step change}, is introduced by Moore \etal \cite{MGBarnabas73} in the context of constructing dendrograms for phylogenetic trees. Subsequently this transformation has been used as a distance function for comparing biological trees \cite{WSmith78}.   Interestingly the authors in \cite{MGBarnabas73} define a local search method based on this operation for finding a locally optimal tree according to a certain biological scoring function. 

Currently the best approximation factor in optimizing the revenue function of Moseley and Wang is due to Alon \etal \cite{AlonAV20}. They get a $0.585$ factor approximation algorithm by a reduction to the Max-Uncut Bisection problem. Currently the best approximation factor for the optimizing Dasgupta's cost function is $O(\sqrt{\log n})$ due to Charikar \etal \cite{CharikarC17} and Cohen-Addad \etal \cite{Cohen-AddadKMM18}.

There are other unsupervised objectives for evaluating hierarchical clusterings. In one example \cite{DasguptaL05}, the goal is to construct a
hierarchical clustering with the guarantee that for every $k \in [n]$ and some $\alpha \ge 1$, the induced $k$-clustering
has cost at most $\alpha$ times the cost of the optimal $k$-clustering. There is a line of work within this framework. See a recent work by  Arutyunova \etal  \cite{ArutyunovaGRSW24} for the latest development on this direction.

%
%
%
%

\section{Analyzing locally optimal trees }
In this section we establish a lower bound for the revenue of the locally optimal trees. 
We begin by introducing a few notations and terminologies. 

For disjoint sets $A$ and $B$, generalizing the notation, let $w(A,B) = \sum_{i\in A, j \in B}w(i,j)$.  The following fact is an immediate consequence of the equations (\ref{def:dasgupta}) and (\ref{def:MW}).

\begin{fact} 
\label{fact:revcost}
For any HC tree $T$ on $n$ leaves, we have $\cost(T)+\rev(T) = n(\sum_{i<j} w(i,j))$. 
\end{fact}

We can define the locally optimal trees using more detailed expressions as stated in the following observation.

\begin{observation}
\label{obs:locallyoptimal} 
$T$ is locally optimal tree if and only if for every edge $(x,y)$ in $T$, as shown in Figure \ref{fig:interchange}, we have 
$$|C|w(A,B) \ge |A|w(B,C)$$
$$|C|w(A,B) \ge |B|w(A,C)$$
\end{observation}

\begin{proof} Consider the trees $T$, $T'$ and $T''$ in the Figure \ref{fig:interchange}. Based on the definition of the revenue function, 
we have $$\rev(T') = \rev(T) +  \Big( |A|w(B,C) - |C| w(A,B) \Big)$$
 $$\rev(T'') = \rev(T) +  \Big( |B|w(A,C) - |C| w(A,B) \Big).$$
Using these equations and  Definition \ref{def:locallyoptimal}, we get the statement made in the observation. 
\end{proof}

Before we delve into the analysis of the local optimal trees, we introduce some notations and definitions that are useful in our arguments. 
Another way to express the cost (or the revenue) function is to spread the cost (revenue) over the merge operations in the tree. 
In each merge operation, two disjoint sets of points, say $A$ and $B$, are merged. Denoting the cost of merging $A$ and $B$ in $T$ by $\mergecost_T(A,B)$, we define 
\begin{equation}
\mergecost_T(A,B) = (|A|+|B|)w(A,B)
\end{equation}
 The following offers an equivalent definition of the cost function. 
\begin{equation}
\cost(T) = \sum_{(A,B) \textrm{ is a merge in } T} \mergecost_T(A,B).
\end{equation}
 In the same manner, for a merge $(A,B)$ in $T$, we define the revenue gained by this merge as
 \begin{equation}  
 \label{eq:mergerev}
 \mergerev_T(A,B) = (|\text{leaves}(T)|-|A|-|B|)w(A,B).
 \end{equation}
  Similarly we have 
 \begin{equation}
 \rev(T) = \sum_{(A,B) \textrm{ is a merge in } T} \mergerev_T(A,B).
 \end{equation}

Now we are ready to state the main theorem.

\begin{theorem}
Let $T$ be a locally optimal tree for a given similarity function $w:[n]\times [n] \rightarrow \mathbb{R}^+$. We have $\rev(T) \ge \frac{n-2}3\sum_{i<j} w(i,j)$.
\end{theorem}

\begin{proof} 
For a HC tree $T$,  let $s(T) = \sum_{i<j, i,j \in \text{leaves}(T)} w(i,j)$.
Given Fact \ref{fact:revcost}, rewriting the statement of the theorem in the new notation, we get $3\rev(T) \ge \cost(T)+ \rev(T) -2 s(T)$. Hence it is enough to show that
\begin{equation}
\label{eqn:thm}
2\rev(T) +2s(T) \ge \cost(T)
\end{equation}
 holds for any locally optimal tree $T$.
 
 We take an inductive approach to prove (\ref{eqn:thm}). As the base case, the statement is clearly true for hierarchical trees with less than $3$ leaves. Now let $T$ be a locally optimal tree with $n > 2$ leaves. Note that every subtree of $T$ is also locally optimal. Let $T_A$ and $T_B$ be the left and right substrees of $T$.  By the induction hypothesis, we have \begin{equation}
 \label{eqn:induction}
2\rev(T_A) +2s(T_A) \ge \cost(T_A), \hspace{1cm} 
 2\rev(T_B) +2s(T_B) \ge \cost(T_B)\end{equation}

Since the subtrees $T_A$ and $T_B$ are joined to create $T$, we can write \begin{equation}\label{eqn:costincrease}\cost(T) = \cost(T_A)+ \cost(T_B) + \mergecost_T(A,B).\end{equation} 
On the other hand, considering the merge revenue as defined in (\ref{eq:mergerev}), we have $\mergerev_T(A,B) = 0$. However the revenue for the merge operations in the subtrees of $T_A$ and $T_B$ is expected to rise. In particular, if $(X,Y)$ is a merge in $T_A$, the revenue of this merge is increased by $|B|w(X,Y)$ as the result of merge $A \cup B$. In other words, we have  $\mergerev_T(X,Y) = \mergerev_{T_A}(X,Y) + |B|w(X,Y)$. Let $\rev^+(T_A,T_B)$ denote the total increase in revenue when $T_A$ and $T_B$ are joined together. Namely,
\begin{equation}
\label{eqn:revincrease}
\rev(T) = \rev(T_A) + \rev(T_B) + \rev^+(T_A,T_B)
\end{equation}   

Note that using the equations (\ref{eqn:induction}), (\ref{eqn:costincrease}) and (\ref{eqn:revincrease}), proving (\ref{eqn:thm}) amounts to showing the following
\begin{equation}
\label{eqn:revcostincrease}
2\rev^+(T_A,T_B) + 2w(A,B) \ge \mergecost_T(A,B) = (|A|+|B|)w(A,B).
\end{equation}  

In order to show (\ref{eqn:revcostincrease}), we prove the following lemma. 

\begin{lemma}
\label{lem:revup}  
Let $F=C \cup D$ be an aribtrary merge in $T_A$ and let $r(F)$ be the increase in revenue in the subtree $T_{F}$ due to the merge of $A$ and $B$. Then $$2r(F) \ge |B|(\frac{|F|-1}{|C|}+\frac{|F|-1}{|D|})w(C,D).$$
\end{lemma}

\begin{proof}
We prove this statement using induction on the number of leaves in $T_{F}$. Consider the situation as depicted in Figure \ref{fig:revup}.  The base case is where $|F|=2$. Here $r(F)=|B|w(C,D)$ which satisfies the statement.

\begin{figure}[H]
\label{fig:revup}
\centering
\begin{tikzpicture}[scale=0.8,transform shape]

\Vertex[x=1.5,y=5.5, size = 0.2, label=$C$, position=left, color=black, fontscale=1.5]{c}
\Vertex[x=3.5,y=6.5, size = 0.2, label=$F$, position=left, color=black, fontscale=1.5]{cd}
\Vertex[x=5.5,y=5.5, size = 0.2, label=$D$, position=right, color=black, fontscale=1.5]{d}

\node[isosceles triangle,
    draw,
    rotate=90,
    fill=white,
    minimum size =1cm] (C1) at (0.5,3.5){};

\Text[x=0.5, y = 2.7]{\large $C_1$}

\node[isosceles triangle,
    draw,
    rotate=90,
    fill=white,
    minimum size =1cm] (C2) at (2.5,3.5){};

\Text[x=2.5, y = 2.7]{\large $C_2$}

\node[isosceles triangle,
    draw,
    rotate=90,
    fill=white,
    minimum size =1cm] (D1) at (4.5,3.5){};
    
\Text[x=4.5, y = 2.7]{\large $D_1$}

\node[isosceles triangle,
    draw,
    rotate=90,
    fill=white,
    minimum size =1cm] (D2) at (6.5,3.5){};
    
\Text[x=6.5, y =2.7]{\large $D_2$}
  
\draw (c) -- (C1.apex);
\draw (cd) -- (c);
\draw (c) -- (C2.apex);
\draw (cd) -- (d);
\draw [dashed, thick] (cd) -- (4.2,7.2);

\draw (d) -- (D2.apex);
\draw (d) -- (D1.apex);

\end{tikzpicture}
\caption{}
\end{figure}

 Now suppose $|C| > 1$ and  $|D|> 1$ where 
 $C$ is created by merging $C_1$ and $C_2$ and
 $D$ is created by merging $D_1$ and $D_2$.  Since $T$ is locally optimal, we have the following set of inequalities.
\begin{equation}
\label{eqn:revup1} |C|w(D_1,D_2) \ge |D_2|w(D_1,C)
\end{equation}
\begin{equation}
\label{eqn:revup2} |C|w(D_1,D_2) \ge |D_1|w(D_2,C) 
\end{equation}
\begin{equation}
\label{eqn:revup3} |D|w(C_1,C_2) \ge |C_2|w(C_1,D) 
\end{equation}
\begin{equation}
\label{eqn:revup4} |D|w(C_1,C_2) \ge |C_1|w(C_2,D) 
\end{equation}

On the other hand, by the induction hypothesis, we have 
\begin{equation}
\label{eqn:ind1}
2r(D) \ge |B|(\frac{|D|-1}{|D_2|}+\frac{|D|-1}{|D_1|})w(D_1,D_2),
\end{equation}  
and,
\begin{equation}
\label{eqn:ind2}
2r(C) \ge |B|(\frac{|C|-1}{|C_2|}+\frac{|C|-1}{|C_1|})w(C_1,C_2)
\end{equation} 

\noindent Multiplying (\ref{eqn:revup1}) by $\frac{|D|-1}{|D_2||C|}$ and (\ref{eqn:revup2}) by $\frac{|D|-1}{|D_1||C|}$ and summing up we get 

\begin{equation}
\label{eqn:d1d2}
(\frac{|D|-1}{|D_2|}+\frac{|D|-1}{|D_1|})w(D_1,D_2) \ge \frac{(|D|-1)}{|C|}w(D,C)
\end{equation}

\noindent Similarly multiplying (\ref{eqn:revup3}) by $\frac{|C|-1}{|C_2||D|}$ and (\ref{eqn:revup4}) by $\frac{|C|-1}{|C_1||D|}$ and summing up we get 

\begin{equation}
\label{eqn:c1c2}
(\frac{|C|-1}{|C_2|}+\frac{|C|-1}{|C_1|})w(C_1,C_2) \ge \frac{(|C|-1)}{|D|}w(D,C)
\end{equation}

The fact that $2r(F) = 2r(C)+2r(D) + 2|B|w(C,D)$, together with (\ref{eqn:ind1}), (\ref{eqn:ind2}), (\ref{eqn:d1d2}), and (\ref{eqn:c1c2}) satisfies the induction claim.
 
The case where one of the sets $C$ and $D$ are of size $1$ is handled similarly. 
Without loss of generality, suppose $|C|=1$ and $D = D_1 \cup D_2$. Here we require $2r(F) \ge |B|(|D|+1)w(C,D)$. From (\ref{eqn:revup1}), (\ref{eqn:revup2}) and (\ref{eqn:ind1}) we get (\ref{eqn:d1d2}). Consequently we get $2r(F) = 2r(D) + 2|B|w(C,D) \ge |B|(D-1)w(C,D) + 2|B|w(B,C) = |B|(|D|+1)w(C,D) $ as desired.

\end{proof}
Note that by symmetry we can write a similar statement for a merge $F=(C\cup D)$ in the subtree $T_B$. Now we use Lemma \ref{lem:revup} to prove (\ref{eqn:revcostincrease}). We claim $2r(A) \ge (|A|-1)w(A,B)$. The case where $|A| =1$ is clearly true. Suppose $A=(A_1\cup A_2)$.  By Lemma \ref{lem:revup}, we have $2r(A)  \ge (\frac{|A|-1}{|A_1|}+\frac{|A|-1}{|A_2|})w(A_1,A_2)$. Since $T$ is locally optimal, we have the inequalities $|B|w(A_1,A_2) \ge |A_1|w(A_2,B)$ and $|B|w(A_1,A_2) \ge |A_2|w(A_1,B)$. Multiplying the former by $\frac{|A|-1}{|A_1|}$ and the latter by $\frac{|A|-1}{|A_2|}$ and summing up we get $2r(A) \ge (|A|-1)w(A,B)$ as claimed. By symmetry, we can show $2r(B) \ge (|B|-1)w(A,B)$. Since $2\rev^+(T_A,T_B) = 2r(A)
+2r(B)$, we get $2\rev^+(T_A,T_B) \ge (|A|+|B|-2)w(A,B)$ which proves   (\ref{eqn:revcostincrease}). This finishes the proof of the theorem. 
\end{proof}


\section{Average Link tree is locally optimal}
In this section, we show the tree computed by the average link algorithm is locally optimal with respect to the interchange operations. The average link algorithm works as follows. Initially each object is put in a separate cluster. Thus initially we have $n$ clusters $C_1,\cdots C_n$. In each step, the algorithm finds the most similar clusters $C_i$ and $C_j$ and merges them into one cluster. The similarity between the clusters $A$ and $B$ is defined as the average similarity between their members, namely $\frac{\sum_{x\in A,y\in B} w(x,y) }{|A||B|}$. The algorithm stops when only one cluster is left. 

\begin{lemma}
\label{lem:AL}
Let $T$ be the hierarchical tree computed by the average link algorithm. $T$ is locally optimal with respect to the interchange operation.  
\end{lemma}

\begin{proof}
Consider the situation of the nodes $x$ and $y$ in the tree $T$ as shown in Figure \ref{fig:interchange} (a). Consider the point where the average link algorithm decides to merge  $A$ and $B$. At this point, the cluster $C$ may not be created yet. Suppose at this point, theree is a collection of disjoint clusters $C_1, \cdots, C_k$ where $C = C_1 \cup \cdots \cup C_k$. The algorithm has decided to merge $A$ and $B$, in particular, because for all $i$, we have
$$sim(B,A) \ge sim(B,C_i)$$
In other words (using $w(A,B) = \sum_{x\in A,y \in B}w(x,y)$)
$$\frac{w(B,A)}{|A||B|} \ge \frac{w(B,C_i)}{|B||C_i|}.$$
Hence
$$|C_i| w(B,A) \ge |A|w(B,C_i).$$
Summing over all $i$, we get
$$\sum_i |C_i| w(B,A) \ge \sum_i |A|w(B,C_i).$$
Since $\sum_i |C_i| = |C|$ and $\sum_i w(B,C_i) = w(B,C)$, we get
$|C| w(B,A) \ge |A|w(B,C).$ 
In the same manner, the fact that $sim(A,B) \ge sim(A,C_i)$ for all $i$, leads to the inequality $ |C| w(B,A) \ge |B|w(A,C)$. These inequalities together imply that performing an interchange on the edge $(x,y)$ does not increase the revenue. Hence based on Observation \ref{obs:locallyoptimal}, $T$ is locally optimal. 
\end{proof}

\section{A distance metric for HC trees}

The local search algorithm presented in this work suggests a distance function for comparing HC trees. 

\begin{definition} 
\label{def:idist}
Let $T_1$ and $T_2$ be two  HC trees.   We define $idist(T_1,T_2)$ as the minimum number of interchange operations needed to convert $T_1$ to $T_2$. 
\end{definition}

The interchange distance between  HC trees is well-defined. In other words, starting from a HC tree $T_1$ of order $n$, we can reach any other  HC tree of order $n$ using only interchange operations. In fact we have the following lemma. 

\begin{lemma}
\label{lem:interchange}
For any two  HC trees $T_1$ and $T_2$ of order $n$, we have $idist(T_1,T_2) = O(n\log n).$
\end{lemma}

Before we give insights on why the above lemma is correct, we show a small consequence of this lemma. In the work by Dasgupta \cite{Dasgupta16}, it has been shown that in the scenario where the underlying similarity function represents a complete unweighted graph (where $w(i,j)=1$ for all pairs $i$ and $j$), any binary tree is deemed optimal (see theorem 3 in \cite{Dasgupta16}). Putting Lemma \ref{lem:interchange} and Observation \ref{obs:locallyoptimal} together, we can present a concise proof of this assertion. Note that for this particular similarity function, any HC tree $T$ is locally optimal. That follows because here for any disjoint pair of sets $A$ and $B$ we have $w(A,B)=|A||B|$. This clearly satisfies the requirement of the being locally optimal according to Observation  \ref{obs:locallyoptimal}. On the other hand, Lemma \ref{lem:interchange} leads to the conclusion that, in this case all locally optimal trees must be globally optimal. 

One can prove Lemma \ref{lem:interchange} in two ways. Let $\mathcal{L}(n)$ denote the set of all (un-rooted) trees with $n$ leaves where each non-leaf node has degree $3$ and each leaf is uniquely labeled by a number from $[n]$. 
 Culik and Wood \cite{CulikW82} showed that the interchange distance over $\mathcal{L}(n)$ is bounded by $4n\log(\frac{n}3) + 4n-12$. 
 By employing a similar approach used in Culik and Wood \cite{CulikW82}, we can establish that the interchange distance over HC trees is bounded by $4n\log n + 4n-12$ (details are left to a full version of this paper.) However, a more precise bound is obtained by considering ordered HC trees.
Here by an {\em ordered HC tree} we mean an HC tree where an ordering is imposed on the children of the nodes. Note that by introducing an ordering on the children, we can interpret the interchange distance using two common operations on binary trees: the rotation and the twist operations. Rotation is frequently utilized in self-organizing binary search trees \cite{AllenM78,Bitner79,SleatorT85}. The twist operation, when applied to node $x$, involves swapping the left and right subtrees of $x$. This operation, as named by Sleator \etal \cite{SleatorTT92}, in combination with rotation, has been used to define a similarity measure for dendrograms \cite{LiZ99}.
 Consider the Figure \ref{fig:interchange}(b). The depicted outcome is a result of a twist operation on node $x$, followed by rotation on node $x$, and then another twist on the same node. The outcome in Figure \ref{fig:interchange}(c) arises from a rotation on $x$, succeeded by a twist on $x$ and then a twist on $y$.

%
Li and Zhang \cite{LiZ99} have defined the twist-rotation distance over ordered  HC trees as the minimum number of twists and rotations needed to convert one ordered  HC tree to another one. They have shown that the twist-rotation distance is bounded by $3n\log n + O(n)$ for ordered HC trees of order $n$. As a consequence, this bound implies that the interchange distance over HC trees of order $n$ is also bounded by $3n\log n + O(n)$.  To see this, as mentioned earlier, an interchange involves a combination of 1 rotation and 2 twist operations. The twist operation does not result in a new tree when the orderings of the children are disregarded. Therefore, we get the claimed bound for the interchange distance as a result.

\section{Implementation of the local search method}
 In this section we present a detailed implementation of the local search method. We consider two variants of the local search method. In one variant, we take a profitable interchange that gives us the maximum amount of increase in revenue. We call this variant local search with {\em greedy interchange}. In the other variant, called local search with {\em random interchange}, we choose a random profitable interchange where the distribution over profitable interchanges is uniform. 

We prove the following lemma for the greedy interchange variant. A similar time and space bound follow for the random interchange variant.

\begin{lemma} Given a HC tree $T$ of order $n$ and the associated similarity function $w:[n]\times [n] \rightarrow \mathbb{R}^+$, there is a data structure for local search with greedy interchange that has the following properties. 
\begin{itemize} 
\item The pre-processing time and space usage of the data structure is $O(n^2)$.
\item There is an algorithm that queries the data structure and finds the most profitable interchange in $O(n)$ time. 
\item The update time of the data structure is $O(n)$. 
\end{itemize}
\end{lemma}

\begin{proof} Let $m$ be the number of nodes in $T$. We have $m = 2n-1$. 
We assume each node $u$ in $T$ is given a unique identifier in $[m]$. In the following, when we speak of a node $u$ in the tree, the value $u$ refers to its unique identifier. For $i \in [m]$, let $T_i$ be the subtree rooted at the node $i$. 
Our data structure is simply a two dimensional array $W$ of dimension $m\times m$ defined as follows.
 
$$W[i][j] =w(\text{leaves}(T_i), \text{leaves}(T_j))$$
We claim the array $W$ is constructed in $O(n^2)$ time. To do this, we first partition the nodes of $T$ into subsets $H_0, \cdots H_t$ where $H_i$ is the set of nodes with height $i$ and $t$ is the height of the tree. Here $H_0$ is the set of leaves in $T$. Such a partitioning can be computed in $O(n)$ time using a bottom-up algorithm for computing the height of every node. Having $H_0, \cdots H_t$, we show how the array $W$ is computed in $O(n^2)$ time in a dynamic programming fashion. First, we compute $W[i][j]$ for every pair $(i,j) \in H_0 \times H_0$. This information is directly available given the simiarlity function $w$. In the subsequent steps, for each  $r = 1, \cdots, t$, we compute $W[i][j]$ for the pairs $(i,j)$ in $H_r \times (H_0 \cup \cdots \cup H_{r-1})$. Note that each entry $W[i][j]$ is computed in $O(1)$ time. To see this, assuming $i$ and $j$ are non-leaf nodes, let $i_1$ and $i_2$ be the left and right children of node $i$. Similarly we define $j_1$ and $j_2$. 
We have $$W[i][j]= W[i_1][j_1]+ W[i_1][j_2]+W[i_2][j_1]+W[i_2][j_2].$$
As the reader can see, the entries required for computing  $W[i][j]$ are already computed in the previous steps. In the case $i$ is childless (a leaf), we have $W[i][j]=W[i][j_1]+W[i][j_2]$. Again the required entries are computed before. Same holds for the case when $j$ is childless. This proves our claim. 

To find the most profitable interchange, we examine every edge $(x,y)$ in the tree (where $x$ is the child of $y$) and calculate the increase in revenue by exchanging the subtree rooted at $z$ (the other child of $y$) with one of the subtrees rooted at the children of $x$ (as shown in Figure \ref{fig:interchange}). By the comments in the proof of the Observation \ref{obs:locallyoptimal} and using the information in the array $W$, the increase in revenue for each interchange is computed in $O(1)$ time. Consequently the best interchange can be computed in $O(n)$ time. 

Now we examine the update time of the data structure. Consider the situation in Figure \ref{fig:interchange}. Suppose the greedy algorithm decides to swap the subtrees $T(B)$ and $T(C)$ which results in tree $T'$. After the swap, the only subtree whose leaves have changed is the subtree rooted at $x$, namely  $T_x$. Therefore to update the two-dimensional array $W$, it is enough to update $W[x][u]$ for all $u \in [m]/\{x\}$. 
Let $z$ and $z'$ be the roots of the subtrees $T(B)$ and $T(C)$ respectively. 
More precisely, for each $u \in [m]/\{x\}$, we set $$W[x][u] \leftarrow W[x][u] + W[z][u] - W[z'][u].$$
Since $m = 2n-1$, the update is done in $O(n)$ time. This finishes the proof of the lemma. 
\end{proof}

\section{Empirical Results}
In this section we present a set of empirical results related to the local search algorithm. For the experiments, we use 4  classical UCI datasets \cite{UCI} as mentioned in Table \ref{table:datasets}. These are relatively small-sized multi-dimensional datasets with varying dimensions.

\begin{table}[H]

\begin{center}
  \begin{tabular}{ | l | c | c | }
    \hline
    Dataset & Size & Dimension  \\ \hline \hline
    Glass & 214 & 10 \\ \hline
    Iris & 150 & 4  \\ \hline
    Zoo & 101 & 16  \\ \hline
    MNIST(*) & 300 & 28$\times$28  \\ \hline
  \end{tabular}
        \caption{The specification of the datasets used in the experiment. (*) A random subset of the MNIST dataset.}
        \label{table:datasets}
\end{center}
\end{table}

For the similarity measure, we have used the Gaussian kernel $K(x,x')=e^{-\frac{\|x-x'\|^2}{2\sigma^2}}$ with $\sigma$ set to the average distance in the dataset. For the first experiment, we have compared two variants of the local search algorithm, the greedy variant (abbreviated as GreedyLS) and the random variant (abbreviated as RandomLS) against the average link algorithm (abbreviated as AL). For each dataset, we have recorded the average and the maximum revenue obtained over the set of $10$ independent runs (the calculated revenue is divided by $ (n-2)\sum_{i<j} w(i,j)$). In each run, we have the begun the the local search from a random tree. 

   \begin{table}[H]
      \begin{center}
  \begin{tabular}{ | c | c |    c |c  |    c  |     c|c  |c | }
    \hline
    Dataset & AL & \multicolumn{3}{c}{GreedyLS} &   \multicolumn{3}{|c|}{RandomLS}  \\ \hline
     &  &  rev|avg & rev|max &  inter\#   & rev|avg & rev|max & inter\#   \\ \hline \hline
    Glass & {\bf 0.5794} &  0.5764 & 0.5785 & 2696   &  0.5788 & 0.5801  & 2068  \\ \hline 
    Iris & {\bf 0.6525} &  0.6488 & 0.6507 & 1200   & 0.6509 & 0.6522 & 1130 \\ \hline 
       Zoo  & 0.6332  & 0.6293 & {\bf 0.6340} & 673    & 0.6317 & 0.6333 & 685  \\ \hline 
       MNIST  & 0.4378 & 0.4406 & {\bf 0.4440} &  3477   & 0.4404 &  0.4430 & 4247  \\ \hline 
  \end{tabular}
  \caption{The results of the first experiment. The columns rev|avg and rev|max show the average and max revenue obtained over the course of 10 independent runs. Here $\uparrow$ is better. The column inter\# shows the number of local steps (interchange operations) taken by the local search algorithm (in average.) Here $\downarrow$ is better. }
    \label{table:exp2}
    \end{center}
\end{table}

The empirical results in Table \ref{table:exp2} shows that results of the local search method are comparable with average link trees in terms of the revenue. In some cases, local search can find trees with higher revenue. However as a drawback, the convergence time of the instances are high. 

In a follow-up experiment, we conducted a comparison of some linkage methods based on the number of local (greedy) steps required to achieve a local optimum. For each scenario, the local search algorithm was initiated from the tree generated by a specific linkage method. The findings presented in Table \ref{table:exp3} suggest that the local search approach can serve as an effective post-processing technique to enhance the revenue generated by the resulting trees.

\begin{table}[H]
      \begin{center}
  \begin{tabular}{ | c |   c |c  |    c |c |    c |c |   }
    \hline
    Dataset &  \multicolumn{2}{c}{single} &  \multicolumn{2}{|c}{complete}  & \multicolumn{2}{|c|}{ward}    \\ \hline
     &  increase (\%) & inter\# &  increase (\%) & inter\#  &  increase (\%) &  inter\#  \\ \hline \hline
    Glass &  1.6 & 128 & 1.6 & 76 &0.9 &{\bf 71} \\ \hline 
    Iris & 1.9  &  115 & 9.3 & {\bf 31} &0.01 &41  \\ \hline 
       Zoo  & 2.0  & 31  &0.71 & {\bf 9} &0.1 &11   \\ \hline 
       MNIST  & 5.4  & 846  &5.3 & {\bf 176} &6.2&201  \\ \hline 
  \end{tabular}
  \caption{Comparison of 3 linkage methods. The column increase \% shows the increased revenue in percentage after the local search is applied. The column inter\# is the number of steps (interchange operations) taken by the greedy local search algorithm.  }
  \label{table:exp3}
    \end{center}
\end{table}

\section{Concluding Remarks }

In this research, we introduced a local search algorithm for hierarchical clustering. The quality of the resulting clustering was evaluated using the objective function proposed by Dasgupta \cite{Dasgupta16}. Specifically, we have demonstrated that the trees generated by the average link exhibit the characteristic of being locally optimal with respect to the interchange operation. Additionally, we have provided an implementation of the local search algorithm. We believe our study can open a new space for further exploration in  local search algorithms for hierarchical clustering. In the following, we highlight some open problems and potential directions for future works.

\begin{itemize}

\item This work leaves open question of the convergence time of the local search method. In particular, can we show that the greedy variant of the local search converges to a local optimum in polynomial time? What about the random variant? For $0,1$ similarity functions, there is a naive upper bound of $O(n^3)$ on the number of local operations. Are there examples where the greedy local search needs $\Omega(n^3)$ or even $\Omega(n^2)$ number of steps to converge? 

\item Another open question is the complexity of computing the interchange distance between two HC trees. In a related problem, Li and Zhang \cite{LiZ99} have shown that computing the twist-rotation distance over HC trees is an NP-hard question. They have also given a polynomial time approximation algorithm with an approximation factor $4\log n + 6$ for this problem. Can we prove a similar result here? 

\item Finally it would be desirable to design faster implementation of the local search method. The $\Theta(n^2)$ space/pre-processing time makes the method impractical for large data sets. For specific similarity functions, such as ones based on the Euclidean distance, can we come up with faster implementations?

\end{itemize}

\bibliographystyle{amsalpha}
\bibliography{references}{}

\newcommand{\etalchar}[1]{$^{#1}$}
\providecommand{\bysame}{\leavevmode\hbox to3em{\hrulefill}\thinspace}
\providecommand{\MR}{\relax\ifhmode\unskip\space\fi MR }
\providecommand{\MRhref}[2]{%
  \href{http://www.ams.org/mathscinet-getitem?mr=#1}{#2}
}
\providecommand{\href}[2]{#2}
\begin{thebibliography}{KMKM17}

\bibitem[AAV20]{AlonAV20}
Noga Alon, Yossi Azar, and Danny Vainstein, \emph{Hierarchical clustering: {A}
  0.585 revenue approximation}, Conference on Learning Theory, {COLT} 2020,
  9-12 July 2020, Virtual Event [Graz, Austria] (Jacob~D. Abernethy and Shivani
  Agarwal, eds.), Proceedings of Machine Learning Research, vol. 125, {PMLR},
  2020, pp.~153--162.

\bibitem[AGK{\etalchar{+}}04]{AryaGKMMP04}
Vijay Arya, Naveen Garg, Rohit Khandekar, Adam Meyerson, Kamesh Munagala, and
  Vinayaka Pandit, \emph{Local search heuristics for k-median and facility
  location problems}, {SIAM} J. Comput. \textbf{33} (2004), no.~3, 544--562.

\bibitem[AGR{\etalchar{+}}24]{ArutyunovaGRSW24}
Anna Arutyunova, Anna Gro{\ss}wendt, Heiko R{\"{o}}glin, Melanie Schmidt, and
  Julian Wargalla, \emph{Upper and lower bounds for complete linkage in general
  metric spaces}, Mach. Learn. \textbf{113} (2024), no.~1, 489--518.

\bibitem[AM78]{AllenM78}
Brian Allen and J.~Ian Munro, \emph{Self-organizing binary search trees}, J.
  {ACM} \textbf{25} (1978), no.~4, 526--535.

\bibitem[BCLP23]{BCLP23}
Lorenzo Beretta, Vincent Cohen{-}Addad, Silvio Lattanzi, and Nikos Parotsidis,
  \emph{Multi-swap k-means++}, Advances in Neural Information Processing
  Systems 36: Annual Conference on Neural Information Processing Systems 2023,
  NeurIPS 2023, New Orleans, LA, USA, December 10 - 16, 2023 (Alice Oh, Tristan
  Naumann, Amir Globerson, Kate Saenko, Moritz Hardt, and Sergey Levine, eds.),
  2023.

\bibitem[Bit79]{Bitner79}
James~R. Bitner, \emph{Heuristics that dynamically organize data structures},
  {SIAM} J. Comput. \textbf{8} (1979), no.~1, 82--110.

\bibitem[CC17]{CharikarC17}
Moses Charikar and Vaggos Chatziafratis, \emph{Approximate hierarchical
  clustering via sparsest cut and spreading metrics}, Proceedings of the
  Twenty-Eighth Annual {ACM-SIAM} Symposium on Discrete Algorithms, {SODA}
  2017, Barcelona, Spain, Hotel Porta Fira, January 16-19 (Philip~N. Klein,
  ed.), {SIAM}, 2017, pp.~841--854.

\bibitem[CCN19]{CharikarCN19}
Moses Charikar, Vaggos Chatziafratis, and Rad Niazadeh, \emph{Hierarchical
  clustering better than average-linkage}, Proceedings of the Thirtieth Annual
  {ACM-SIAM} Symposium on Discrete Algorithms, {SODA} 2019, San Diego,
  California, USA, January 6-9, 2019 (Timothy~M. Chan, ed.), {SIAM}, 2019,
  pp.~2291--2304.

\bibitem[CGH{\etalchar{+}}22]{Cohen-AddadGHOS22}
Vincent Cohen{-}Addad, Anupam Gupta, Lunjia Hu, Hoon Oh, and David Saulpic,
  \emph{An improved local search algorithm for k-median}, Proceedings of the
  2022 {ACM-SIAM} Symposium on Discrete Algorithms, {SODA} 2022, Virtual
  Conference / Alexandria, VA, USA, January 9 - 12, 2022 (Joseph~(Seffi) Naor
  and Niv Buchbinder, eds.), {SIAM}, 2022, pp.~1556--1612.

\bibitem[CGPR20]{ChooGPR20}
Davin Choo, Christoph Grunau, Julian Portmann, and V{\'{a}}clav Rozhon,
  \emph{k-means++: few more steps yield constant approximation}, Proceedings of
  the 37th International Conference on Machine Learning, {ICML} 2020, 13-18
  July 2020, Virtual Event, Proceedings of Machine Learning Research, vol. 119,
  {PMLR}, 2020, pp.~1909--1917.

\bibitem[CKM19]{Cohen-AddadKM19}
Vincent Cohen{-}Addad, Philip~N. Klein, and Claire Mathieu, \emph{Local search
  yields approximation schemes for k-means and k-median in euclidean and
  minor-free metrics}, {SIAM} J. Comput. \textbf{48} (2019), no.~2, 644--667.

\bibitem[CKMM18]{Cohen-AddadKMM18}
Vincent Cohen{-}Addad, Varun Kanade, Frederik Mallmann{-}Trenn, and Claire
  Mathieu, \emph{Hierarchical clustering: Objective functions and algorithms},
  Proceedings of the Twenty-Ninth Annual {ACM-SIAM} Symposium on Discrete
  Algorithms, {SODA} 2018, New Orleans, LA, USA, January 7-10, 2018 (Artur
  Czumaj, ed.), {SIAM}, 2018, pp.~378--397.

\bibitem[CYL{\etalchar{+}}20]{ChatziafratisYL20}
Vaggos Chatziafratis, Grigory Yaroslavtsev, Euiwoong Lee, Konstantin
  Makarychev, Sara Ahmadian, Alessandro Epasto, and Mohammad Mahdian,
  \emph{Bisect and conquer: Hierarchical clustering via max-uncut bisection},
  The 23rd International Conference on Artificial Intelligence and Statistics,
  {AISTATS} 2020, 26-28 August 2020, Online [Palermo, Sicily, Italy] (Silvia
  Chiappa and Roberto Calandra, eds.), Proceedings of Machine Learning
  Research, vol. 108, {PMLR}, 2020, pp.~3121--3132.

\bibitem[Das16]{Dasgupta16}
Sanjoy Dasgupta, \emph{A cost function for similarity-based hierarchical
  clustering}, Proceedings of the 48th Annual {ACM} {SIGACT} Symposium on
  Theory of Computing, {STOC} 2016, Cambridge, MA, USA, June 18-21, 2016
  (Daniel Wichs and Yishay Mansour, eds.), {ACM}, 2016, pp.~118--127.

\bibitem[DHL{\etalchar{+}}16]{DasGuptaH0TZ16}
Bhaskar DasGupta, Xin He, Ming Li, John Tromp, and Louxin Zhang, \emph{Nearest
  neighbor interchange and related distances}, Encyclopedia of Algorithms
  (2016), 1402--1405.

\bibitem[DL05]{DasguptaL05}
Sanjoy Dasgupta and Philip~M. Long, \emph{Performance guarantees for
  hierarchical clustering}, J. Comput. Syst. Sci. \textbf{70} (2005), no.~4,
  555--569.

\bibitem[FRS19]{FriggstadRS19}
Zachary Friggstad, Mohsen Rezapour, and Mohammad~R. Salavatipour, \emph{Local
  search yields a {PTAS} for k-means in doubling metrics}, {SIAM} J. Comput.
  \textbf{48} (2019), no.~2, 452--480.

\bibitem[GMB73]{MGBarnabas73}
M.~Goodman G.W.~Moore and J.~Barnabas, \emph{An iterative approach from the
  standpoint of the additive hypothesis to the dendrogram problem posed by
  molecular data sets}, J. Theoret. Biology \textbf{38} (1973), 423--457.

\bibitem[HG05]{HellerG05}
Katherine~A. Heller and Zoubin Ghahramani, \emph{Bayesian hierarchical
  clustering}, Machine Learning, Proceedings of the Twenty-Second International
  Conference {(ICML} 2005), Bonn, Germany, August 7-11, 2005 (Luc~De Raedt and
  Stefan Wrobel, eds.), {ACM} International Conference Proceeding Series, vol.
  119, {ACM}, 2005, pp.~297--304.

\bibitem[IW82]{CulikW82}
Karel~Cul{\'{\i}}k II and Derick Wood, \emph{A note on some tree similarity
  measures}, Inf. Process. Lett. \textbf{15} (1982), no.~1, 39--42.

\bibitem[KMKM17]{KobrenMKM17}
Ari Kobren, Nicholas Monath, Akshay Krishnamurthy, and Andrew McCallum, \emph{A
  hierarchical algorithm for extreme clustering}, Proceedings of the 23rd {ACM}
  {SIGKDD} International Conference on Knowledge Discovery and Data Mining,
  Halifax, NS, Canada, August 13 - 17, 2017, {ACM}, 2017, pp.~255--264.

\bibitem[KMN{\etalchar{+}}04]{KanungoMNPSW04}
Tapas Kanungo, David~M. Mount, Nathan~S. Netanyahu, Christine~D. Piatko, Ruth
  Silverman, and Angela~Y. Wu, \emph{A local search approximation algorithm for
  k-means clustering}, Comput. Geom. \textbf{28} (2004), no.~2-3, 89--112.

\bibitem[LS19]{LattanziS19}
Silvio Lattanzi and Christian Sohler, \emph{A better k-means++ algorithm via
  local search}, Proceedings of the 36th International Conference on Machine
  Learning, {ICML} 2019, 9-15 June 2019, Long Beach, California, {USA}
  (Kamalika Chaudhuri and Ruslan Salakhutdinov, eds.), Proceedings of Machine
  Learning Research, vol.~97, {PMLR}, 2019, pp.~3662--3671.

\bibitem[LZ99]{LiZ99}
Ming Li and Louxin Zhang, \emph{Twist-rotation transformations of binary trees
  and arithmetic expressions}, J. Algorithms \textbf{32} (1999), no.~2,
  155--166.

\bibitem[MC12]{MurtaghC12}
Fionn Murtagh and Pedro Contreras, \emph{Algorithms for hierarchical
  clustering: an overview}, WIREs Data Mining Knowl. Discov. \textbf{2} (2012),
  no.~1, 86--97.

\bibitem[MKK{\etalchar{+}}19]{MonathKKGM19}
Nicholas Monath, Ari Kobren, Akshay Krishnamurthy, Michael~R. Glass, and Andrew
  McCallum, \emph{Scalable hierarchical clustering with tree grafting},
  Proceedings of the 25th {ACM} {SIGKDD} International Conference on Knowledge
  Discovery {\&} Data Mining, {KDD} 2019, Anchorage, AK, USA, August 4-8, 2019
  (Ankur Teredesai, Vipin Kumar, Ying Li, R{\'{o}}mer Rosales, Evimaria Terzi,
  and George Karypis, eds.), {ACM}, 2019, pp.~1438--1448.

\bibitem[MLZ96]{LiTZ96}
J.~Tromp M.~Li and L.~Zhang, \emph{On the nearest neighbour interchange
  distance between evolutionary trees}, J. Theoret. Biology \textbf{182}
  (1996), 436--467.

\bibitem[MRS{\etalchar{+}}19]{MenonRSCCK19}
Aditya~Krishna Menon, Anand Rajagopalan, Baris Sumengen, Gui Citovsky, Qin Cao,
  and Sanjiv Kumar, \emph{Online hierarchical clustering approximations}, CoRR
  \textbf{abs/1909.09667} (2019).

\bibitem[MW17]{MoseleyW17}
Benjamin Moseley and Joshua~R. Wang, \emph{Approximation bounds for
  hierarchical clustering: Average linkage, bisecting k-means, and local
  search}, Advances in Neural Information Processing Systems 30: Annual
  Conference on Neural Information Processing Systems 2017, December 4-9, 2017,
  Long Beach, CA, {USA} (Isabelle Guyon, Ulrike von Luxburg, Samy Bengio,
  Hanna~M. Wallach, Rob Fergus, S.~V.~N. Vishwanathan, and Roman Garnett,
  eds.), 2017, pp.~3094--3103.

\bibitem[ST85]{SleatorT85}
Daniel~Dominic Sleator and Robert~Endre Tarjan, \emph{Self-adjusting binary
  search trees}, J. {ACM} \textbf{32} (1985), no.~3, 652--686.

\bibitem[STT92]{SleatorTT92}
Daniel~Dominic Sleator, Robert~Endre Tarjan, and William~P. Thurston,
  \emph{Short encodings of evolving structures}, {SIAM} J. Discret. Math.
  \textbf{5} (1992), no.~3, 428--450.

\bibitem[UCI]{UCI}
\emph{{UCI datasets}, howpublished = {\url{https://archive.ics.uci.edu/}}, note
  = {Accessed: 2024-05-07}}.

\bibitem[WS78]{WSmith78}
M.S. Waterman and T.F. Smith, \emph{On the similarity of dendograms}, J.
  Theoret. Biology \textbf{75} (1978), 789--800.

\end{thebibliography}

\end{document}